%% file: hepro4_lenain.tex
\documentclass{ws-ijmpcs}

\usepackage{overcite}
\usepackage{hyperref}
\usepackage{xspace}
\newcommand\hess{H.E.S.S.\xspace}%
\newcommand\fermi{\textit{Fermi}/LAT\xspace}%
\newcommand\g{\ensuremath{\gamma}}%

\input{aas_macros.tex}

\begin{document}

\markboth{Jean-Philippe Lenain, for the \hess collaboration}
{The \hess extragalactic sky}

\catchline{}{}{}{}{}
%
%%%%%%%%%%%%%%%%%%%%%%%%%%%%%%%%%%%%%%%%%%%%%%%%%%%%%%%%%%%%%%%%%%%%

\title{THE \hess EXTRAGALACTIC SKY}

\author{JEAN-PHILIPPE LENAIN, FOR THE \hess COLLABORATION}

\address{LPNHE, Universit\'e Pierre et Marie Curie Paris 6, Universit\'e Denis Diderot Paris 7, CNRS/IN2P3, 4 Place Jussieu\\
  75252 Paris Cedex 5, France.\\
jlenain@lpnhe.in2p3.fr}

\maketitle

\begin{history}
\received{Day Month Year}
\revised{Day Month Year}
\end{history}

\begin{abstract}
More than fifty extragalactic very high energy (VHE; $E>100$\,GeV) sources have been found using ground-based imaging atmospheric Cherenkov telescopes, about twenty of which have been discovered using the \hess (High Energy Stereoscopic System) experiment based in Namibia. Even though BL\,Lac objects are the dominant class of VHE detected extragalactic objects, other types of sources (starburst galaxies, radio galaxies or flat spectrum radio quasars) begin to emerge. A review of the extragalactic sources studied with \hess is given, with an emphasis on new results.
\end{abstract}

\ccode{PACS numbers: 95.85.Pw, 98.54.Cm, 98.54.Gr, 98.65.-r, 98.70.Rz, 98.70.Vc}

\section{The \hess Experiment}

\begin{figure}[pt]
\centering
\includegraphics[width=0.9\textwidth]{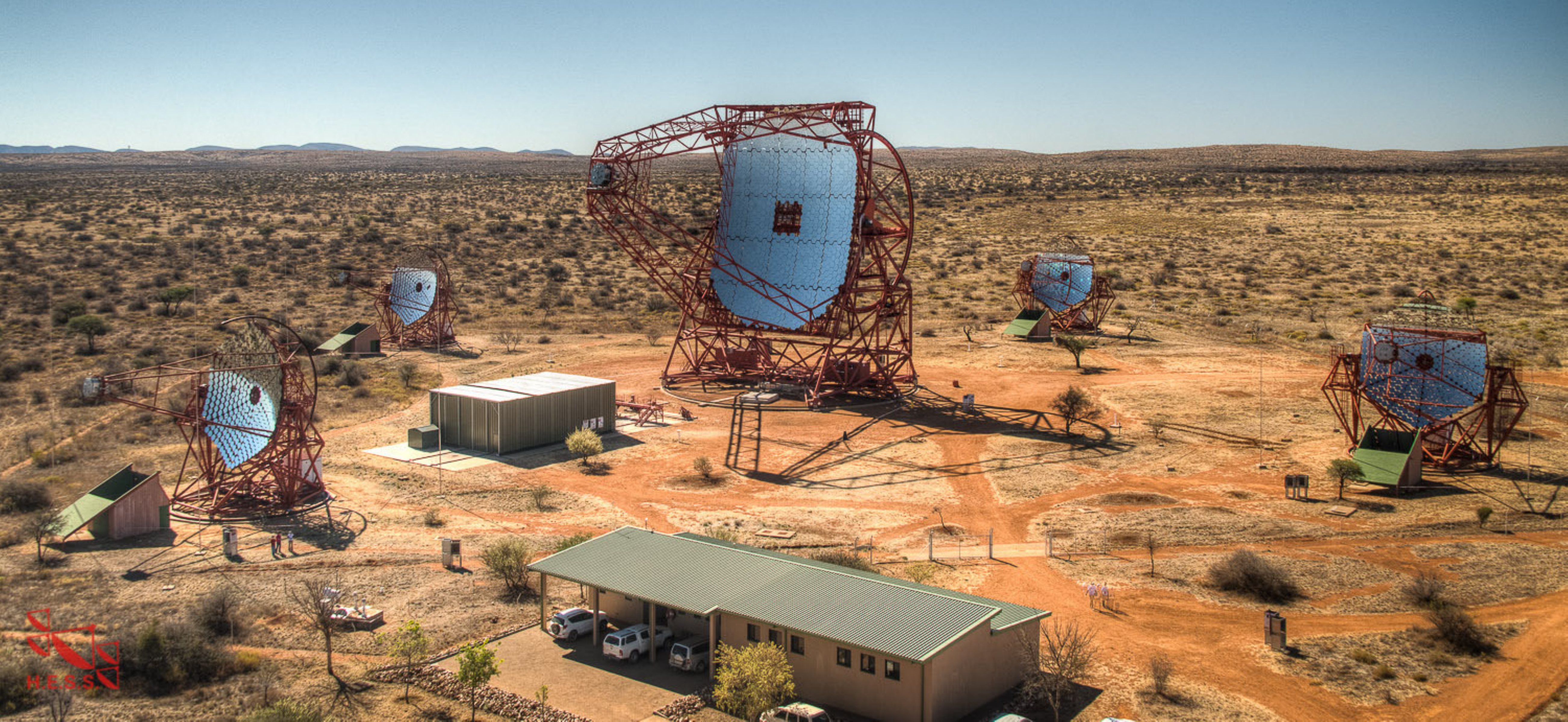}
\caption{The \hess array of five imaging atmospheric Cherenkov telescopes in Namibia.}
\label{fig:hess}
\end{figure}

% HESS and HESS II
The \hess experiment has been used for more than ten years to explore the very high energy (VHE; $E>100$\,GeV) sky in the Southern hemisphere with four imaging atmospheric Cherenkov telescopes (IACT) operated in Namibia. In September 2012, a fifth telescope located in the middle of the original array was inaugurated (see Fig.~\ref{fig:hess}). Its mirror surface of 600\,m$^2$ will enable a notable gain in sensitivity as well as a lower energy threshold of a few tens of GeV.

% Studied objects with H.E.S.S.
\hess is used to study different physics topics, such as the origin of Galactic cosmic rays, the acceleration processes at work in compact objects, or searching for indirect signs of dark matter in the Universe. Observations of many types of objects are thus performed with \hess, from supernova remnants to galaxy clusters, pulsar wind nebul\ae\ or active galactic nuclei (AGN). This manuscript focuses on recent results obtained with \hess on extragalactic sources.

% TeVCat
The number of known very high energy sources has dramatically increased with the advent of the current generation of IACT, with now almost 150 detected objects. At the time of writing this manuscript, from the 58 detected extragalactic VHE emitters, 52 objects are of the blazar type with 49 BL\,Lac objects\footnote{See TeVCat at \href{http://tevcat.uchicago.edu}{http://tevcat.uchicago.edu}.}.

\section{Recent \hess Results}

\subsection{Starburst galaxies}

\begin{figure}
\centering
\includegraphics[width=0.9\textwidth]{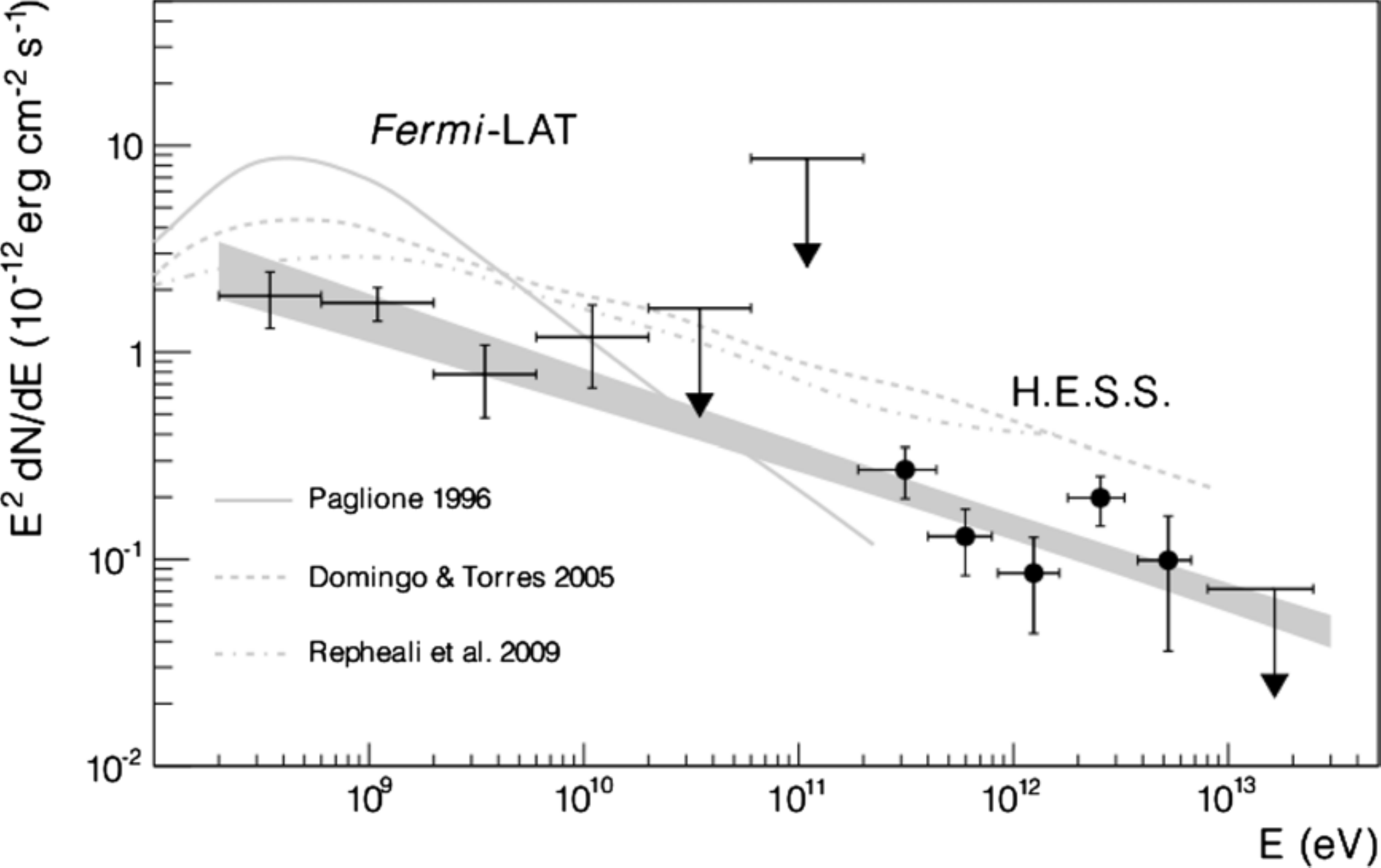}
\caption{Spectral energy distribution of NGC\,253 observed with \hess and \fermi, with data fit simultaneously in shaded area (1$\sigma$ confidence band) and theoretical predictions. See Ref.~\protect\refcite{2012ApJ...757..158A} for more details.}
\label{fig:ngc253}
\end{figure}

Following deep observations of more than 170\,h of good quality data, the starburst galaxy NGC\,253 was detected at a 7.1$\sigma$ confidence level (CL) and a spectrum with a photon index of $\Gamma=2.14 \pm0.18_\mathrm{stat}$ was derived\cite{2012ApJ...757..158A}. Detailed analysis of both \hess and public \fermi data revealed a joint spectrum with a photon index of $\Gamma=2.34 \pm 0.03$ in the high and very high energy ranges (see Fig.~\ref{fig:ngc253}). Studying this high energy emission, compared to theoretical models, the \g-ray emission is likely to be dominated by hadronic interactions within the starburst region. More than 20\% of the non-thermal energy is converted into $\pi^0$ \g-ray production, assuming an efficiency of 10\% for cosmic ray acceleration processes from supernova remnants in the starburst region.

\subsection{Radio galaxies}

Up to now, three radio galaxies have been detected in the VHE domain, namely NGC\,1275 \cite{2012A+A...539L...2A}, M\,87 \cite{2003A+A...403L...1A,2006Sci...314.1424A} and Cen\,A \cite{2009ApJ...695L..40A}, all of them belonging to the FR-I type. The VHE emitting source IC\,310 also used to be classified as a head-tail radio galaxy, even though recent studies have revealed blazar-like properties\cite{2011ApJ...743..171A,2013arXiv1305.5147T}.

More specifically, the radio galaxy M\,87 has been observed for more than 10 years at high energies, along with extended multi-wavelength campaigns in the X-ray, optical and radio domains. Three major flares were detected at VHE \cite{2006Sci...314.1424A,2009Sci...325..444A,2012ApJ...746..151A}, in 2005, 2008 and 2010. However, no unique multi-wavelength pattern was identified in this object. Indeed the question of the location of the emission zone arose when the 2005 flare showed multi-wavelength correlation between observations in VHE and X-ray ranges from the knot {\it HST}-1, while the temporal analysis of the two other flares seemed to establish the central part of the jet as the high energy emitter\cite{2009Sci...325..444A,2012ApJ...746..151A}. In 2010, the flare profile was unambiguously observed as asymmetrical, with a characteristic variability timescale of the order of one day. In radio, VLBA 43\,GHz observations show no indication for an enhanced flux from the base of the jet in the 2010 event, contrasting with radio observations of the 2008 flare for which a radio burst was detected contemporaneously with VHE \g-ray emission. The complex variability behaviour of this source particularly strengthens the case for strictly simultaneous multi-wavelength observations to help pinpointing the emission processes at work in high energy emitting objects.

Cen\,A, the closest FR\,I radio galaxy, has been detected at VHE with \hess\cite{2009ApJ...695L..40A} using more than 120\,h of observations. The \fermi collaboration reported the detections of its giant radio lobes\cite{2010Sci...328..725A} as well as central AGN activity at high energies\cite{2010ApJ...719.1433A}. Refs.~\refcite{2009ApJ...695L..40A} and~\refcite{2010ApJ...719.1433A} have shown that the high and VHE spectra are only marginally compatible, and simple one-zone synchrotron self-Compton (SSC) modelling applied to this object requires an unusually low Doppler factor, with respect to BL\,Lac objects, due to the large angle of the jet axis to the line of sight\cite{2010ApJ...719.1433A}. These facts support the need for more elaborate interpretations, such as multi-zone models\cite{2000A+A...358..104C,2008A+A...478..111L}, decelerating flows\cite{2003ApJ...594L..27G} or high energy emission processes in the vicinity of the central black hole magnetosphere\cite{2009A+A...506L..41R}. More recently, a potential additional hard spectral component was found in \fermi data\cite{2013ApJ...770L...6S}, emerging at around 4\,GeV and reconciling the marginal mismatch between \fermi and \hess data, further supporting the presence of multiple emitting components. The observations of time variability at high energies, not detected so far, would definitely help in distinguishing between the currently viable scenarios to explain this complex source.

\subsection{Gamma-ray bursts}

Gamma-ray bursts show evidence of strongly enhanced relativistic outflows and are established as powerful particle accelerators up to the high energy domain\cite{2013arXiv1303.2908F}. However, so far no \g-ray burst has been detected in the VHE domain, even though extensive search has been followed on this path. Both the enhanced sensitivity and probed energy range with the future Cherenkov Telescope Array (CTA) promise the detection of about a few events per year\cite{2013APh....43..134M,2013APh....43..252I}.

Serendipitous observations of GRB\,060602B were made with the \hess array, when an alert was received for an event occurring in the field of view of \hess observations at that time, resulting in observations of both the prompt and afterglow phases at VHE\cite{2009ApJ...690.1068A}. The \hess observations did not unveil any significant VHE \g-ray signal, but the very soft \textit{Swift}/BAT spectrum and its proximity to the Galactic centre led Ref.~\refcite{2009MNRAS.393..126W} to favor a Galactic X-ray burster origin for this peculiar event, instead of a cosmological one.

\subsection{Galaxy clusters}

Powerful active galactic nuclei can be found at the centre of some galaxy clusters, whose outbursts generate feedback reactions in the embedding intra-cluster medium. Such interactions can be detected for instance by the presence of cavities in the X-ray emission\cite{2007ARA+A..45..117M}. Cosmic rays represent a viable solution to explain the main pressure support of such bubbles. In this case, \g-ray emission is expected in these objects: inelastic collisions of cosmic rays with the thermal ambient medium would result in VHE \g-rays from hadronic processes, while electronic cosmic rays would inverse Compton up-scatter with the cosmic microwave background or the extragalactic background light.

Hydra\,A is the closest galaxy cluster hosting such an AGN activity. The \textit{Chandra} satellite revealed a large scale cavity built up from three generations of AGN outbursts\cite{2007ApJ...659.1153W}. Both giant outer lobes, dominating the system energetics, and inner lobes, filled with the youngest population of particles, are of interest for the search for \g-ray emission. \fermi and \hess observations of Hydra\,A resulted in no detection in the high and very high energy domains\cite{2012A+A...545A.103H}. However, the upper limits derived on the \g-ray flux help constraining theoretical models: for the outer giant lobes, in which hadronic processes dominate, a full mixing of cosmic rays within the intra-cluster medium is excluded, with a degree of mixing less than 0.7 and 0.5, respectively from \hess and \fermi data. In the case of leptonic emission originating from the inner lobes, \fermi and \textit{XMM} data tend to favour a high magnetic field strength above 8\,$\mu$G.

\subsection{Flat spectrum radio quasars}

\begin{figure}
\centering
\includegraphics[width=0.7\textwidth]{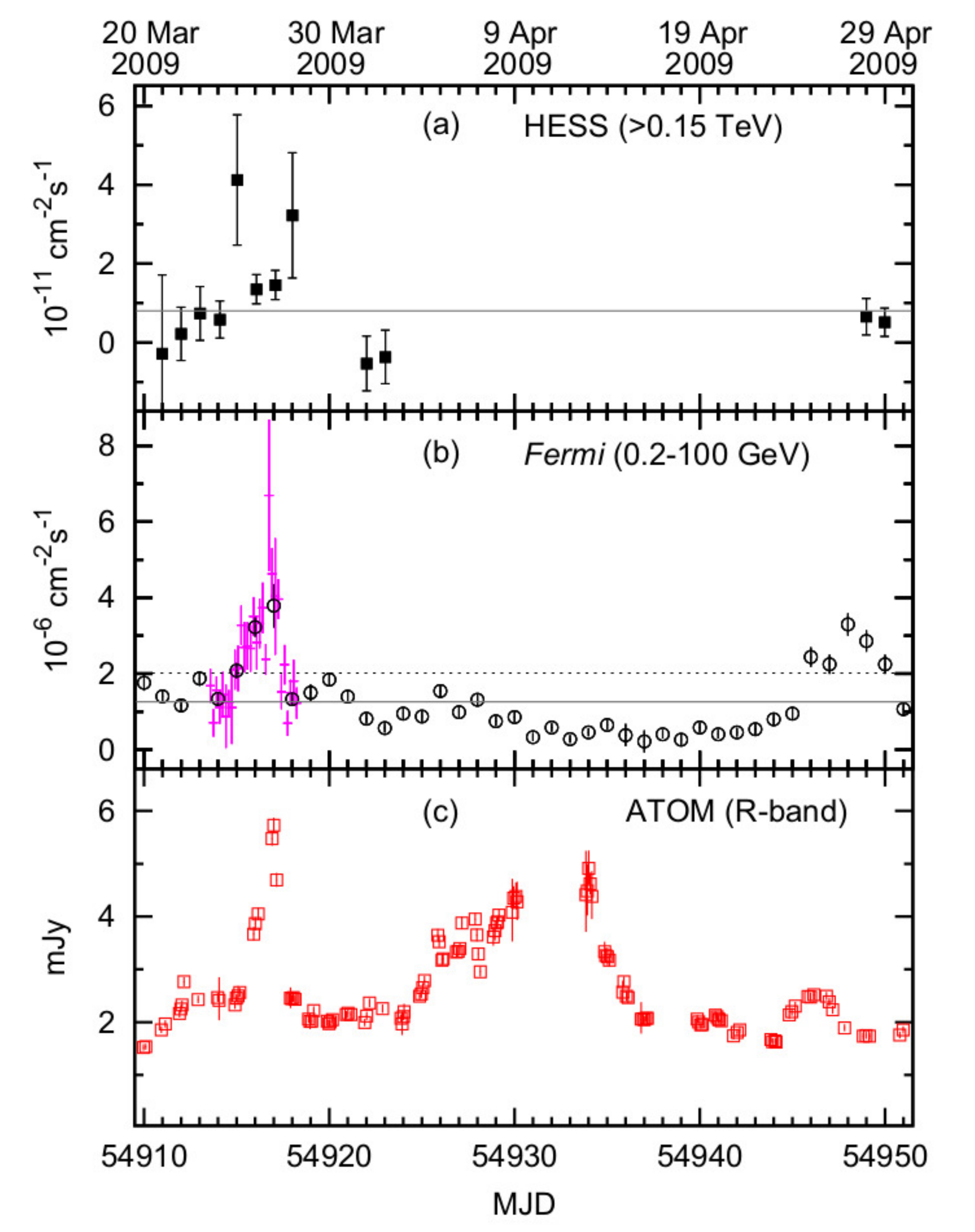}
\caption{Light curve of PKS\,1510$-$089 in the VHE, high energy, and optical ranges. See Ref.~\protect\refcite{2013A+A...554A.107H} for more details.}
\label{fig:pks1510}
\end{figure}

Due to the synchrotron peak of their emission lying at low frequency and their soft \g-ray spectrum, flat spectrum radio quasars (FSRQ) are rare at VHE, with only three known sources: 4C\,$+$21.35\cite{2011ApJ...730L...8A}, 3C\,279\cite{2008Sci...320.1752M} and PKS\,1510$-$089\cite{2013A+A...554A.107H}. The later was detected at a 9.2$\sigma$ CL in about 16\,h of good quality \hess data. Even though no significant variability was detected in the VHE range, contemporaneous observations with \fermi in the high energies and ATOM in the optical band revealed strong variability on a day scale (see Fig.~\ref{fig:pks1510}). The VHE spectrum was used to derive upper limits on the photon density of the extragalactic background light (EBL, see Section~\ref{sec:ebl}). For this particularly far -- for VHE standards -- object ($z=0.361$), these limits were found to be compatible to those derived from relatively nearby BL\,Lac objects\cite{2006Natur.440.1018A}.

\subsection{BL\,Lac objects}

BL\,Lac objects are by far the most numerous known extragalactic VHE emitters. This can be explained by the fact that their jet, pointing close to the observer's line of sight, Doppler boost their broadband emission, with the higher frequency-peaked objects being natural candidates for \g-ray emission\cite{1998MNRAS.299..433F}.

\begin{figure}
\centering
\includegraphics[width=0.9\textwidth]{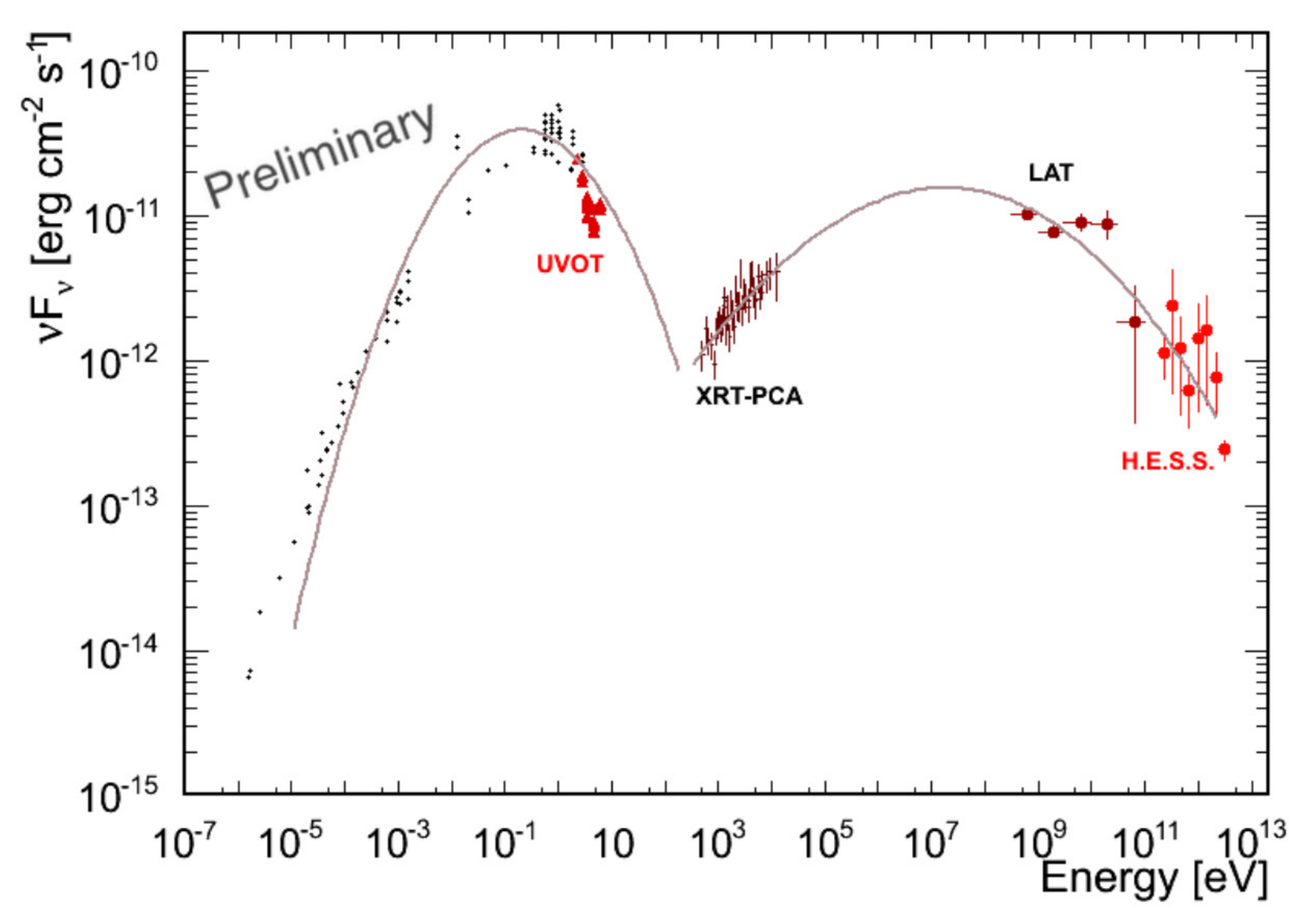}
\caption{Spectral energy distribution of AP\,Lib. See Ref.~\protect\refcite{2010tsra.confE.199F} for more details.}
\label{fig:aplib}
\end{figure}

Among these, the high-frequency-peaked BL\,Lac object (HBL) PKS\,0447$-$439 was discovered at VHE with \hess at the 15.2$\sigma$ CL in 13.5\,h of observations\cite{2013A+A...552A.118H}. Multi-wavelength observations with \fermi, \textit{Swift}, ATOM and ROTSE were used to draw a contemporaneous spectral energy distribution of this source, which was modelled using an SSC model. The redshift of this object being uncertain\cite{1997AJ....114.1356C,1998AJ....115.1253P,2012MNRAS.423L..84L,2012AIPC.1505..566P}, the modelling of multi-wavelength data was used to derive an upper limit on its redshift of $z<0.59$.

1ES\,1312$-$423 is an HBL which was serendipitously discovered\cite{2013MNRAS.434.1889H} with \hess in the field of view of observations of Cen\,A. Correcting for the large offset acceptance, the source is detected at the 5.7$\sigma$ CL in 48.4\,h. 1ES\,1312$-$423 was found to be very faint in the VHE range, with a flux of $\sim0.5\%$ of the Crab flux, making it one the faintest VHE source currently known. Contemporaneous multi-wavelength data from radio to high energies with ATCA, ATOM, \textit{Swift} and \fermi were used to study its spectral energy distribution, which can be well described using a one-zone SSC model.

AP\,Lib, a low-frequency-peaked BL\,Lac object, was discovered as a VHE \g-ray emitter with \hess at 7$\sigma$ CL in 11\,h\cite{2010tsra.confE.199F}. From \fermi data analysis, it was found that this object also exhibits a hard \g-ray spectrum at high energy with $\Gamma=2.1 \pm 0.1$ in the LAT energy range. Together with contemporaneous \textit{Swift} and \textit{RXTE} data, the spectral energy distribution of AP\,Lib displays an atypical, unusually broad component at high energy, with respect to other VHE emitters (see Fig.~\ref{fig:aplib}). Such a component seriously challenges simple SSC scenarios, and multiple zone models\cite{2008A+A...478..111L,2008MNRAS.385L..98T}, external inverse Compton\cite{1985MNRAS.212..553S} or hadronic processes\cite{1991A+A...251..723M,2000NewA....5..377A,2001APh....15..121M} may have to be invoked to interpret its multi-wavelength emission.

\subsection{Extragalactic background light}
\label{sec:ebl}

\begin{figure}
\centering
\includegraphics[width=0.9\textwidth]{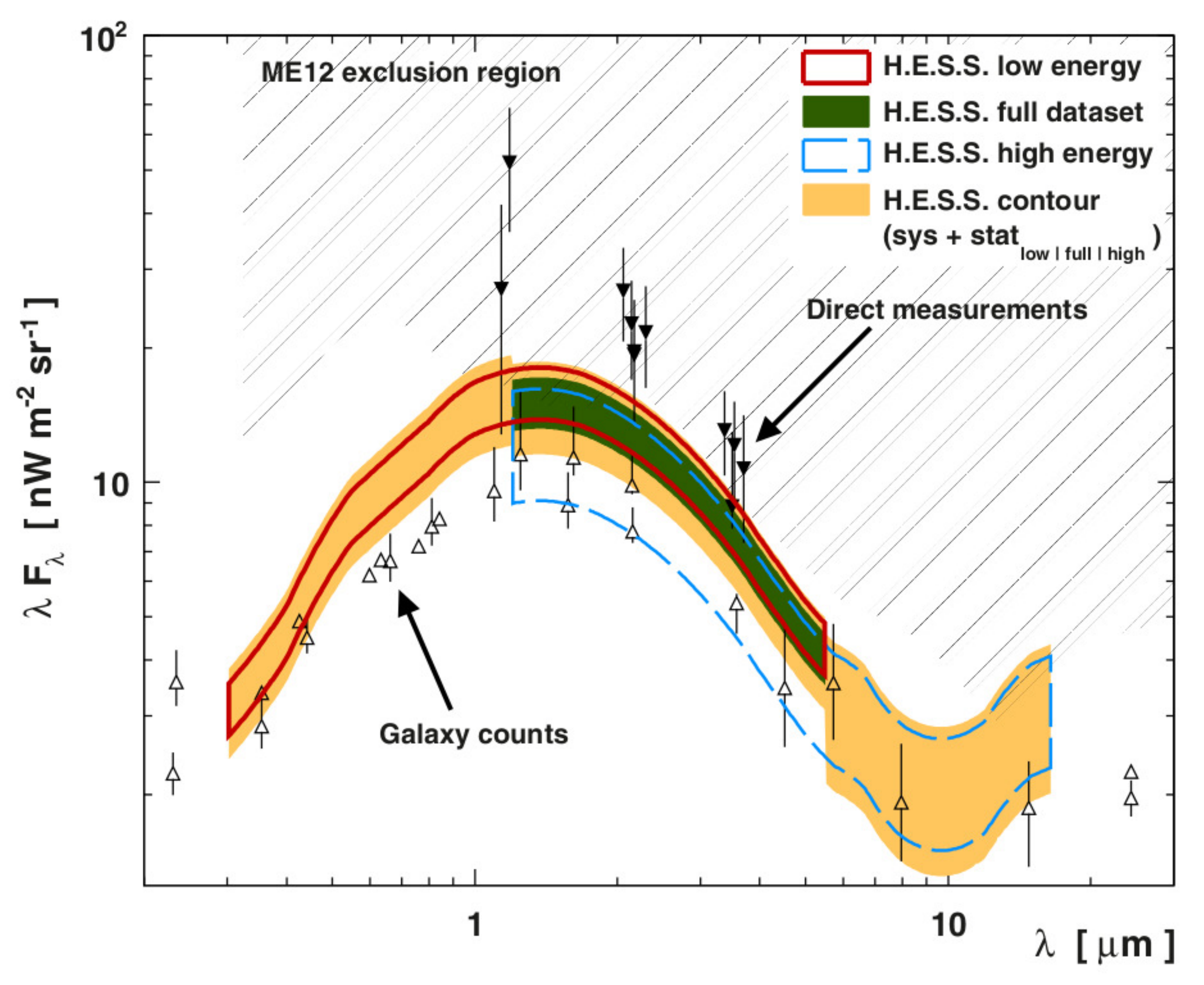}
\caption{Detection of the EBL with \hess See Ref.~\protect\refcite{2013A+A...550A...4H} for more details.}
\label{fig:ebl}
\end{figure}

The properties of the extragalactic background light\cite{2001ARA+A..39..249H} (EBL), the sum of the light emitted by stars and galaxies and its reprocessing by dust, have long been studied both using direct -- in the infrared and optical bands -- and indirect measurements. The VHE photons emitted in a source interact with the EBL field, softening the intrinsic VHE spectrum as observed from the Earth. By accumulating statistics from the brightest blazars detected with \hess, the global imprint of the EBL in VHE \g-ray observations has been significantly revealed at the 8.8$\sigma$ CL\cite{2013A+A...550A...4H} (see Fig.~\ref{fig:ebl}). The \fermi collaboration also detected such an imprint at high energies for more distant blazars\cite{2012Sci...338.1190A}, up to $z\simeq 1.6$.

\section{Conclusions}

The recent \hess results on VHE extragalactic sources show a broad diversity, with e.g. flux upper limits obtained both in the prompt and afterglow phases of GRB\,060602B, constraining limits put on galaxy cluster emission models, as well as the need for simultaneous multi-wavelength studies of AGN to better understand the physics at the sources, and last but not least the detection of the EBL imprint on blazar spectra.

With the ramp-up of the fifth \hess telescope, more results are expected to come soon, both in monoscopic mode with events triggering the biggest telescope alone, as well as hybrid events that trigger it and at least another telescope in the array. Those heterogeneous data taking capabilities will ensure both an energy threshold as low as possible, and a better sensitivity in the overall energy range. In this regard, distant cosmological sources could be observed with \hess, to better understand both the intrinsic emission at work and further investigation of properties of the EBL. The energy coverage will also help investigating the radiative processes in AGN, and specifically FSRQ, by probing the high energy peak of their spectra. In a further future, the CTA\cite{2013APh....43....3A} will help widening our views on the VHE emitting Universe.

\section*{Acknowledgments}

I would like to thank the conference organisers for the opportunity to speak at the \textit{4$^\mathit{th}$ High Energy Phenomena in Relativistic Outflows} (HEPRO IV) meeting, and for organising an interesting and enjoyable conference.

The support of the Namibian authorities and of the University of Namibia in facilitating the construction and operation of \hess is gratefully acknowledged, as is the support by the German Ministry for Education and Research (BMBF), the Max Planck Society, the German Research Foundation (DFG), the French Ministry for Research, the CNRS-IN2P3 and the Astroparticle Interdisciplinary Programme of the CNRS, the UK Science and Technology Facilities Council (STFC), the IPNP of the Charles University, the Czech Science Foundation, the Polish Ministry of Science and Higher Education, the South African Department of Science and Technology and National Research Foundation, and by the University of Namibia. We appreciate the excellent work of the technical support staff in Berlin, Durham, Hamburg, Heidelberg, Palaiseau, Paris, Saclay, and in Namibia in the construction and operation of the equipment.

\bibliographystyle{ws-ijmpcs}
\bibliography{hepro4_lenain.bbl}

\end{document}

%% file: aas_macros.tex
\def\jnl@style{\itshape}
\def\aaref@jnl#1{{\jnl@style#1}}
\def\aaref@jnl#1{{\jnl@style#1}}

\def\aj{\aaref@jnl{AJ}}                   % Astronomical Journal
\def\araa{\aaref@jnl{ARA\&A}}             % Annual Review of Astron and Astrophys
\def\apj{\aaref@jnl{ApJ}}                 % Astrophysical Journal
\def\apjl{\aaref@jnl{ApJ}}                % Astrophysical Journal, Letters
\def\apjs{\aaref@jnl{ApJS}}               % Astrophysical Journal, Supplement
\def\ao{\aaref@jnl{Appl.~Opt.}}           % Applied Optics
\def\apss{\aaref@jnl{Ap\&SS}}             % Astrophysics and Space Science
\def\aap{\aaref@jnl{A\&A}}                % Astronomy and Astrophysics
\def\aapr{\aaref@jnl{A\&A~Rev.}}          % Astronomy and Astrophysics Reviews
\def\aaps{\aaref@jnl{A\&AS}}              % Astronomy and Astrophysics, Supplement
\def\azh{\aaref@jnl{AZh}}                 % Astronomicheskii Zhurnal
\def\baas{\aaref@jnl{BAAS}}               % Bulletin of the AAS
\def\jrasc{\aaref@jnl{JRASC}}             % Journal of the RAS of Canada
\def\memras{\aaref@jnl{MmRAS}}            % Memoirs of the RAS
\def\mnras{\aaref@jnl{MNRAS}}             % Monthly Notices of the RAS
\def\pra{\aaref@jnl{Phys.~Rev.~A}}        % Physical Review A: General Physics
\def\prb{\aaref@jnl{Phys.~Rev.~B}}        % Physical Review B: Solid State
\def\prc{\aaref@jnl{Phys.~Rev.~C}}        % Physical Review C
\def\prd{\aaref@jnl{Phys.~Rev.~D}}        % Physical Review D
\def\pre{\aaref@jnl{Phys.~Rev.~E}}        % Physical Review E
\def\prl{\aaref@jnl{Phys.~Rev.~Lett.}}    % Physical Review Letters
\def\pasp{\aaref@jnl{PASP}}               % Publications of the ASP
\def\pasj{\aaref@jnl{PASJ}}               % Publications of the ASJ
\def\qjras{\aaref@jnl{QJRAS}}             % Quarterly Journal of the RAS
\def\skytel{\aaref@jnl{S\&T}}             % Sky and Telescope
\def\solphys{\aaref@jnl{Sol.~Phys.}}      % Solar Physics
\def\sovast{\aaref@jnl{Soviet~Ast.}}      % Soviet Astronomy
\def\ssr{\aaref@jnl{Space~Sci.~Rev.}}     % Space Science Reviews
\def\zap{\aaref@jnl{ZAp}}                 % Zeitschrift fuer Astrophysik
\def\nat{\aaref@jnl{Nature}}              % Nature
\def\iaucirc{\aaref@jnl{IAU~Circ.}}       % IAU Cirulars
\def\aplett{\aaref@jnl{Astrophys.~Lett.}} % Astrophysics Letters
\def\apspr{\aaref@jnl{Astrophys.~Space~Phys.~Res.}}
\def\bain{\aaref@jnl{Bull.~Astron.~Inst.~Netherlands}} 
\def\fcp{\aaref@jnl{Fund.~Cosmic~Phys.}}  % Fundamental Cosmic Physics
\def\gca{\aaref@jnl{Geochim.~Cosmochim.~Acta}}   % Geochimica Cosmochimica Acta
\def\grl{\aaref@jnl{Geophys.~Res.~Lett.}} % Geophysics Research Letters
\def\jcp{\aaref@jnl{J.~Chem.~Phys.}}      % Journal of Chemical Physics
\def\jgr{\aaref@jnl{J.~Geophys.~Res.}}    % Journal of Geophysics Research
\def\jqsrt{\aaref@jnl{J.~Quant.~Spec.~Radiat.~Transf.}}
\def\memsai{\aaref@jnl{Mem.~Soc.~Astron.~Italiana}}
\def\nphysa{\aaref@jnl{Nucl.~Phys.~A}}   % Nuclear Physics A
\def\physrep{\aaref@jnl{Phys.~Rep.}}   % Physics Reports
\def\physscr{\aaref@jnl{Phys.~Scr}}   % Physica Scripta
\def\planss{\aaref@jnl{Planet.~Space~Sci.}}   % Planetary Space Science
\def\procspie{\aaref@jnl{Proc.~SPIE}}   % Proceedings of the SPIE